
\magnification \magstep1
\raggedbottom
\openup 2\jot
\voffset6truemm
\headline={\ifnum\pageno=1\hfill\else
\hfill{\it Linear Form of Canonical Gravity}
\hfill \fi}
\centerline {\bf LINEAR FORM OF CANONICAL GRAVITY}
\vskip 1cm
\centerline {GIAMPIERO ESPOSITO and COSIMO STORNAIOLO}
\vskip 0.3cm
\noindent
{\it Istituto Nazionale di Fisica Nucleare,
Sezione di Napoli, Mostra d'Oltremare Padiglione 20,
80125 Napoli, Italy;}
\vskip 0.3cm
\noindent
{\it Dipartimento di Scienze Fisiche, Mostra d'Oltremare
Padiglione 19, 80125 Napoli, Italy.}
\vskip 0.3cm
\noindent
{\bf Summary.} - Recent work in the literature has shown that
general relativity can be formulated in terms of a jet bundle
which, in local coordinates, has five entries: local coordinates
on Lorentzian space-time, tetrads, connection one-forms,
multivelocities corresponding to the tetrads and multivelocities
corresponding to the connection one-forms. The derivatives of
the Lagrangian with respect to the latter class of multivelocities
give rise to a set of multimomenta which naturally occur in
the constraint equations. Interestingly, all the constraint
equations of general relativity are linear in terms of this class
of multimomenta. This construction has been then extended to
complex general relativity, where Lorentzian space-time is
replaced by a four-complex-dimensional complex-Riemannian
manifold. One then finds a holomorphic theory where the familiar
constraint equations are replaced by a set of equations linear
in the holomorphic multimomenta, providing such multimomenta
vanish on a family of two-complex-dimensional surfaces.
In quantum gravity, the problem arises to quantize a real
or a holomorphic theory on the extended space where the
multimomenta can be defined.
\vskip 100cm
Recent work by the authors [1] has shown that a constraint
analysis of general relativity can be performed starting from
a one-jet bundle which, in local coordinates, is represented
by $\Bigr(x^{a},e_{\; \; {\hat b}}^{a},
\omega_{a}^{\; \; {\hat b}{\hat c}},V_{\; \; b{\hat c}}^{a},
W_{ab}^{\; \; \; {\hat c}{\hat d}}\Bigr)$. With our notation,
$x^{a}$ are local coordinates on space-time,
$e_{\; \; {\hat b}}^{a}$ are the tetrad vectors,
$\omega_{a}^{\; \; {\hat b}{\hat c}}$ the connection one-forms,
$V_{\; \; b {\hat c}}^{a}$ the multivelocities corresponding to
$e_{\; \; {\hat b}}^{a}$ and $W_{ab}^{\; \; \; {\hat c}{\hat d}}$
the multivelocities corresponding to
$\omega_{a}^{\; \; {\hat b}{\hat c}}$. In analogy with classical
mechanics, where the momenta are the derivatives of the
Lagrangian with respect to the velocities, one can here define
the {\it multimomenta} (cf. ref. [1])
$$
{\widetilde \pi}_{a}^{\; \; b{\hat c}} \equiv
{\partial L \over \partial V_{\; \; b{\hat c}}^{a}}
\; ,
\eqno (1)
$$
and
$$
{\tilde p}_{\; \; \; {\hat c}{\hat d}}^{ab} \equiv
{\partial L \over \partial W_{ab}^{\; \; \; {\hat c}{\hat d}}}
\; .
\eqno (2)
$$
In general relativity, the former vanish, while the latter
take the form
$$
{\tilde p}_{\; \; \; {\hat c}{\hat d}}^{ab}
={e\over 2}\Bigr(e_{\; \; {\hat c}}^{a} \;
e_{\; \; {\hat d}}^{b}
-e_{\; \; {\hat d}}^{a} \; e_{\; \; {\hat c}}^{b}\Bigr)
\; ,
\eqno (3)
$$
where $e$ is the determinant of the tetrad. Note that the
analogy with classical mechanics is more strict, if one
thinks that in both cases one is working with a {\it finite}
number of degrees of freedom [2]. However, since in field theory
the base space is the whole space-time manifold, and jet
bundles have as base space a fibre bundle over space-time,
one has to introduce the multimomenta instead of the momenta.
This leads to a fully covariant formalism [3].

The main result of
ref. [1] was that, on any spacelike hypersurface $\Sigma$, the
constraint equations are all linear in the multimomenta (3),
and take the form (see ref. [1] for the notation)
$$
\int_{\Sigma}\lambda^{{\hat c}{\hat d}}
\Bigr(D_{a}{\tilde p}^{ab}\Bigr)_{{\hat c}{\hat d}}
d^{3}x_{b}=0
\; ,
\eqno (4)
$$
$$
\int_{\Sigma}{\rm Tr}\biggr[{\tilde p}^{af}\Omega_{ad}
-{1\over 2}{\tilde p}^{ab}\Omega_{ab} \; \delta_{d}^{f}
\biggr]u^{d} \; d^{3}x_{f}=0
\; ,
\eqno (5)
$$
providing the multimomenta (3) vanish on the boundary
$\partial \Sigma$ of $\Sigma$. The alternative possibility
is to set to zero at the boundary the gauge parameters
$\lambda^{{\hat c}{\hat d}}$, or the connection one-forms
$\omega_{a}^{\; \; {\hat b}{\hat c}}$, jointly with the vector
field $u^{a}$ describing diffeomorphisms on space-time [1].

In complex general relativity, the five entries of the
corresponding jet bundle (see above) are all of holomorphic
nature [4], and it is incorrect to talk about a Cauchy
problem, since the concepts of spacelike hypersurface and
time evolution are meaningless. The basic {\it postulate}
of our multisymplectic approach is instead that, on
evaluating the holomorphic multimomentum map [4] on an
arbitrary three-complex-dimensional surface $\Sigma_{c}$,
and setting the resulting geometric object to zero, one gets
all the basic equations of the theory, which correspond to
the constraint equations of the Lorentzian theory. They are
the holomorphic counterpart of these constraints, but cannot
quite be called constraints themselves. The explicit
calculation [4] shows indeed that, providing the holomorphic
multimomenta vanish on the two-complex-dimensional boundary
$\partial \Sigma_{c}$ of $\Sigma_{c}$, the holomorphic
equations still take the forms (4) and (5), where $\Sigma$
is replaced by $\Sigma_{c}$, and all geometric objects have
now a holomorphic nature [4].

The implications for twistor theory are discussed in ref. [4].
We are instead interested in the relevance of the
multisymplectic scheme for quantum gravity. The basic problems
and properties can be described as follows.
\vskip 0.3cm
\noindent
(i) How to define suitable brackets for a space-time covariant
analysis.
\vskip 0.3cm
\noindent
(ii) How to build the counterpart of Dirac's map and Dirac's
quantization scheme.
\vskip 0.3cm
\noindent
(iii) What are the arguments of the state vectors, and how to
define a Feynman path integral with our Lagrangian [1].
\vskip 0.3cm
\noindent
(iv) Our holomorphic framework yields a theory {\it without time},
and the Lorentzian theory cannot be recovered by imposing
suitable reality conditions. In our complex base space the complex
conjugation of spinors is not invariant under holomorphic coordinate
transformations [5], and hence cannot be defined. The unprimed
and primed spin-spaces become then independent of each other,
not related by any conjugation [5]. By contrast, in the Ashtekar
programme [6], one studies complex tetrads on a
four-real-dimensional Lorentzian manifold. This hybrid scheme
makes it possible, in principle, to recover real general
relativity. However, it does not make full use of the holomorphic
formalism (the base space remaining real Lorentzian), and it can
be quite hard to find a suitable set of reality conditions.
\vskip 0.3cm
\noindent
(v) If a complex Ricci-flat space-time, not necessarily
anti-self-dual, could be reconstructed out of a twistor
space consisting of charges for massless
spin-3/2 fields as suggested by
Penrose (see chapter 5 of ref. [5] and references therein),
one might try to relate the quantization of the classical
multisymplectic formalism to the quantization of the twistor
scheme.

These exciting problems are being investigated for the
first time within the multimomenta formulation. The elegance
of the mathematical formalism seems to suggest that one
has to learn how to formulate physical laws and quantization
schemes on the extended spaces where multimomenta can be
defined, instead of the original spaces, where the
constraint equations are quadratic in the momenta [6].
The results obtained in refs. [1,4], and the quantization
programme discussed in this note, seem to point out that yet
a new formalism is available for canonical gravity and
the non-perturbative quantization of the gravitational
field. Whether or not this can improve the current theories
of classical and quantum gravity, will depend on a deeper
understanding of the mathematical structures underlying
Lagrangian field theory and spinor geometry.
\vskip 1cm
\leftline {REFERENCES}
\vskip 1cm
\item {[1]}
ESPOSITO G., GIONTI G. and STORNAIOLO C., {\it Nuovo Cimento B},
{\bf 110} (1995). [GR-QC 9506008]
\item {[2]}
KIJOWSKI J., {\it Commun. Math. Phys.}, {\bf 30} (1973) 99.
\item {[3]}
GOTAY M. J., A Multisymplectic Framework for Classical Field
Theory and the Calculus of Variations I: Covariant Hamiltonian
Formalism, in: M. Francaviglia, ed., {\it Mechanics, Analysis
and Geometry: 200 Years After Lagrange} (North Holland,
Amsterdam, 1991) p. 203.
\item {[4]}
ESPOSITO G. and STORNAIOLO C., {\it Class. Quantum Grav.},
{\bf 12} (1995) 1733.
\item {[5]}
ESPOSITO G., {\it Complex General Relativity, Fundamental
Theories of Physics}, Vol. 69 (Kluwer, Dordrecht) 1995.
\item {[6]}
ASHTEKAR A., {\it Lectures on Non-Perturbative Canonical
Gravity} (World Scientific, Singapore) 1991.
\bye